\documentclass[manuscript,screen,nonacm]{acmart}
\AtBeginDocument{%
  }

\setcopyright{acmlicensed}
\copyrightyear{2026}
\acmYear{2026}
\acmDOI{XXXXXXX.XXXXXXX}
\acmJournal{JETC}

\setcopyright{none}

\usepackage{multirow}

\usepackage{amsmath}

\usepackage{url}

\usepackage{colortbl}
\usepackage{hhline}
\usepackage{relsize}

\usepackage{pifont}
\usepackage{marvosym}
\usepackage{ragged2e}

\newcommand{\ignore}[1]{ }
\usepackage[dvipsnames,svgnames]{xcolor}
\usepackage{pifont}
\usepackage{marvosym}
\usepackage{ragged2e}
\usepackage{multirow}
\usepackage{vcell}

\usepackage{courier}

\usepackage[ruled,vlined]{algorithm2e}

\usepackage{multirow}
\usepackage{bm} 

\newcommand{\red}{}

\renewcommand{\red}{\textcolor{red}}

\usepackage{soul}

\usepackage{setspace}

\usepackage[table]{xcolor}

\begin{document}

\title{Maximizing Memory-Level Parallelism via Integrated Stochastic Logic-in-Memory Architectures}

\author{Farzad Razi}
\orcid{0000-0001-6930-4553}
\authornote{Both authors contributed equally to this research.}
\email{frazi@umn.edu}
\affiliation{%
  \institution{University of Minnesota}
  \city{Minneapolis}
  \state{MN}
  \country{USA}
}

\author{Mehran Moghadam}
\orcid{0000-0002-1325-1664}
\email{moghadam@case.edu}
\authornotemark[1]
\affiliation{%
  \institution{Case Western Reserve University}
  \city{Cleveland}
  \state{OH}
  \country{USA}}

\author{Sercan Aygun}
\orcid{0000-0002-4615-7914}
\email{sercan.aygun@louisiana.edu}
\affiliation{%
  \institution{University of Louisiana at Lafayette}
  \city{Lafayette}
  \state{LA}
  \country{USA}
}

\author{M. Hassan Najafi}
\orcid{0000-0002-4655-6229}
\email{najafi@case.edu}
\affiliation{%
 \institution{Case Western Reserve University}
 \city{Cleveland}
 \state{OH}
 \country{USA}}

\author{Marc Riedel}
\orcid{0000-0002-3318-346X}
\email{mriedel@umn.edu}
\affiliation{%
  \institution{University of Minnesota}
  \city{Minneapolis}
  \state{MN}
  \country{USA}}

\renewcommand{\shortauthors}{Razi et al.}

\begin{abstract}
Today's high-performance architectures are increasingly constrained by 
data movement latency and energy overhead, as the slowdown of single-core performance scaling coincides with the rise of highly data-intensive workloads. 
In-memory architectures have emerged as a complementary solution to conventional von Neumann systems 
by alleviating memory bandwidth bottlenecks, exploiting massive concurrency, and mitigating excessive data movement between memory and processing units. 
This study proposes a parallel in-memory stochastic computing (SC) architecture that implements an end-to-end computation pipeline within Magnetic Tunnel Junction (MTJ)-based memory augmented with logic-in-memory (LIM) capabilities. By leveraging the inherent stochasticity and write-read characteristics of MTJ devices, the proposed architecture enables a fully parallel and deterministic conversion of binary operands into probabilistic bit-streams, eliminating the need for energy-intensive external random number generation circuitry.
These bit-streams are processed by parallel stochastic arithmetic units integrated directly within the memory arrays to efficiently implement core arithmetic and transcendental functions with minimal hardware complexity and inherent noise tolerance. The resulting stochastic outputs can be either reused as an input of future stochastic processing or 
converted back to binary form using parallel accumulation mechanisms and stored in the MTJ memory. By tightly integrating data storage, bit-stream generation, and computation within a unified in-memory fabric, the proposed design maximizes memory-level parallelism while substantially minimizing data movement. Experimental evaluations demonstrate that the proposed architecture achieves a $22\times$--$64\times$ speedup compared to serial stochastic implementations and reduces computational latency by over three orders of magnitude compared to state-of-the-art in-memory computing approaches, while exhibiting superior fault tolerance under noise levels as high as 30\%.
\end{abstract}




\maketitle

\section{Introduction}
\label{Intro}

Parallel computing is a central organizing principle in modern hardware architectures, enabling systems to deliver high performance and energy efficiency under increasingly stringent power and scalability constraints~\cite{6307773}. As traditional single-core performance scaling has diminished, architectural innovation has shifted toward exploiting parallelism across multiple hardware layers, including multi-core processors, vector and \textit{Single Instruction, Multiple Data} (SIMD) units, and memory-level parallelism~\cite{10.1145/1128022.1128023}. This approach enables concurrent execution of operations and more effective utilization of both on-chip and off-chip resources, particularly for workloads characterized by large data volumes and repetitive computation. Application domains such as scientific computing~\cite{10.1137/070693199}, data analytics~\cite{10.1145/1327452.1327492}, and image processing~\cite{Vasile2024ImageProcessingHardwareAcceleration} exemplify these trends, as they often involve structured, data-parallel operations that can be executed independently across large data sets. By explicitly exposing and efficiently managing parallelism at the hardware level, modern architectures can reduce execution latency, improve throughput, and achieve better energy proportionality, establishing parallel computing a key enabler for contemporary high-performance and distributed systems.

In-memory computing (IMC) has emerged as an architectural approach to address the growing performance and energy challenges associated with data movement in modern computing systems~\cite{5695526}. In conventional von Neumann architectures, frequent transfers of data between memory and processing units incur significant latency and energy overhead, particularly for data-intensive workloads. By enabling computation to be performed directly within or near memory arrays, IMC significantly reduces unnecessary data transfers and improves data locality~\cite{Sebastian2020MemoryDevicesInMemoryComputing}. Recent advances in emerging memory technologies, such as magnetic tunnel junctions (MTJs), further reinforce this paradigm by offering non-volatility, high endurance, and tight integration of storage and computation~\cite{Jung2022, Lv2024MTJCRAM}. While IMC primarily targets data-movement inefficiencies, 
parallel computing addresses throughput limitations by enabling concurrent execution across multiple processing elements. When combined, these paradigms address complementary bottlenecks related to locality and serialization, reducing memory bandwidth pressure, data movement energy overhead, and pipeline serialization while improving the utilization of inherent memory-level parallelism. Together, they provide a scalable foundation for achieving high performance and energy efficiency in modern data-intensive systems.

IMC architectures naturally favor lightweight and regular circuit structures, as complex control logic and wide datapaths can diminish the benefits of reduced data movement. In this context, stochastic computing (SC)~\cite{WeikangSurvey, Gaines} offers an attractive alternative to conventional complex binary radix architectures by representing numerical values as probabilistic bit-streams, where information is encoded in the ratio of \textit{ones} to \textit{zeros} rather than in fixed-position binary formats. This encoding enables arithmetic operations to be implemented using extremely simple logic primitives, while allowing accuracy to be traded for latency or energy by adjusting the bit-stream length. Consequently, operations such as multiplication and accumulation can be realized with minimal hardware, while offering inherent tolerance to noise and variation, properties that are particularly advantageous in memory-centric computing environments~\cite{10.1145/3240765.3240797}. This paradigm is particularly well suited to MTJ-based memories, as MTJ devices can be leveraged to efficiently generate stochastic bit-streams, a longstanding bottleneck in practical SC systems. By leveraging the inherent randomness and write-read behavior of MTJs, 
random or pseudo-random bit-streams can be generated in situ without relying on complex or energy-intensive external random number generation circuitry. 

In this work, we propose a parallel in-memory SC architecture that integrates data storage, stochastic conversion, and computation into an end-to-end computation pipeline within MTJ-based memory augmented with logic-in-memory (LIM) capabilities.
The proposed architecture begins by reading binary operands from MTJ memory arrays, followed by a deterministic and fully parallel conversion of the binary values into stochastic bit-streams. This conversion exploits 
MTJ \textit{write} and \textit{read} operations across memory rows, eliminating the need for external random number generators. The generated bit-streams are then processed by parallel stochastic arithmetic units implemented directly within memory, enabling concurrent execution of core arithmetic operations. Finally, the stochastic results are converted back into binary form using parallel accumulation mechanisms and written back to the MTJ memory. By tightly integrating data storage, stochastic conversion, computation, and write-back within a unified parallel architecture, the proposed design enables efficient in-memory execution of complex operations, while minimizing data movement and maximizing memory-level parallelism. We evaluate the proposed approach using representative image processing workloads, demonstrating its effectiveness in improving both performance and energy efficiency for data-intensive applications.
The remainder of this paper is organized as follows. Section~\ref{background} provides a background in SC and MTJ structures. Section~\ref{prop} presents the proposed parallel in-memory architecture. Section~\ref{exp} evaluates the accuracy, robustness, latency, and power efficiency of the proposed architecture together with the related IMC works comparison and parallel image processing case study. Finally, Section~\ref{conc} concludes the paper.

\begin{figure*}[!t]
  \centering
  \includegraphics[width=0.6\textwidth]{Figures/SC_BS.pdf}
  \caption{Conventional SC bit-stream generation approach using RNG and comparator (a) Uncorrelated bit-stream using independent RNGs, (b) Correlated bit-streams by sharing an RNG unit, (c) Serial bit-streams generated in multiple cycles, and (d) Parallel (bundle) bit-streams generated in a single cycle.}
  \label{fig:sc_bs}
  \vspace{-0.75em}
\end{figure*}

\section{Background 
}
\label{background}
\subsection{Stochastic Computing (SC)}
SC has emerged as an alternative to conventional binary computing,
offering massively parallel and low-cost structures for implementing complex arithmetic operations owing to its simple arithmetics with few logic gates~\cite{alaghi_survey2013}. By trading 
precision for hardware simplicity and fault tolerance, SC enables a pathway for deploying complex algorithms on resource-constrained hardware where traditional binary logic becomes impractical.
In SC, real numbers are represented using random bit-streams, with unipolar encoding for values in the 
$[0,1]$ interval and bipolar encoding for values in the $[-1,1]$ interval. The encoded values are determined by the probability of observing a logic ‘1’ over the length of the bit-stream~\cite{6361409}. 
The key advantage of SC is in realizing complex arithmetic operations using simple standard logic primitives (e.g., \texttt{AND}/\texttt{XNOR} gates for multiplication and multiplexer 
for scaled addition). Computational accuracy can be tuned by adjusting the length of bit-streams ($N$) and can be affected by the randomness quality of the input bit-streams.
The simplicity and high parallelism of SC are particularly attractive for IMC fabrics, where regular, lightweight compute structures scale better than wide datapaths and complex control. Fig.~\ref{fig:sc_bs}(a) and (b) demonstrate the common approach to generate \textit{uncorrelated} and \textit{correlated} SC bit-streams using independent and shared random number generator (RNG) sources, respectively. In both cases, each bit of the stochastic stream is produced by comparing a random number against a target constant number. 
Fig.~\ref{fig:sc_bs}(c) and (d) further show that stochastic bit-streams can be generated and processed either serially or in parallel, enabling a trade-off between latency (i.e., proportional to the bit-stream length) and hardware cost.

\begin{figure}[!t]
  \centering
  \includegraphics[width=225pt]{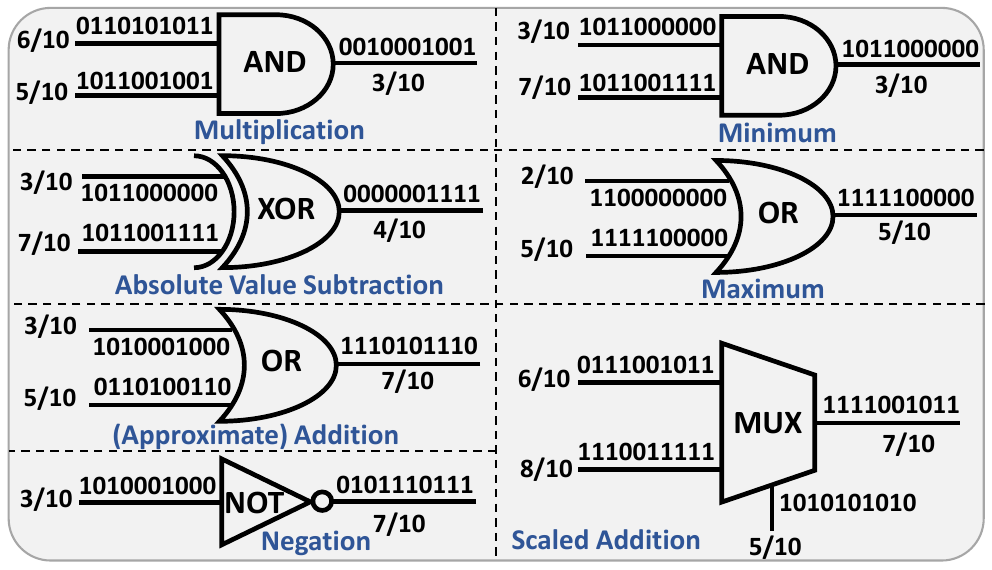}
  \vspace{-1em}
  \caption{Basic arithmetic operations using simple logic gates in SC. Some arithmetic operations demand uncorrelated input bit-streams (e.g., \textit{multiplication} and \textit{approx. addition}), while others require correlated ones (e.g., \textit{minimum} and \textit{maximum}) to obtain accurate results.
  }
  \label{fig:sc_op}
  \vspace{-1.0em}
\end{figure}

A central challenge in practical SC lies in efficient bit-stream generation and correlation management. State-of-the-art approaches utilize quasi-random or low-discrepancy sequences to produce
high-quality, uncorrelated bit-streams~\cite{Alaghi_Fast,Siting_Parallel_Sobol_TVLSI2018,Najafi_TVLSI_2019,p2lsg-ASP-DAC24}. 
To preserve SC's low hardware cost, most prior designs rely on serial processing, which fundamentally limits throughput. In contrast, this work targets memory-level parallel SC by generating large bundles of stochastic bit-streams directly within the memory arrays. By exploiting the massive internal parallelism of memory operations, the proposed approach processes many bits concurrently. 
This transforms SC from a sequential, bit-serial bottleneck into a throughput-oriented, array-parallel execution model that naturally aligns with the strengths of LIM architectures.

SC is also well known for its inherent fault tolerance. Because information is distributed across many time/space samples, transient 
errors, device variability, and soft faults tend to perturb only a fraction of bits in the stream, translating to a gradual degradation in the estimated value rather than a catastrophic failure of specific significance bits as in conventional binary arithmetic. When coupled with in-memory execution, this property enables robust computation under non-ideal sensing/write behaviors and noisy device physics: the architecture can trade stream length (or parallel accumulation depth) for accuracy, while maintaining high throughput by leveraging parallelism across memory rows/columns. These characteristics directly motivate the goal of this work: a fully parallel, end-to-end in-memory SC pipeline that (i) maximizes memory-level parallelism for stochastic arithmetic and transcendental operators and (ii) preserves accuracy under noise and faults through SC's statistical representation, as we will discuss in Section~\ref{exp}.
Fig.~\ref{fig:sc_op} illustrates basic SC arithmetic operations, including multiplication, scaled and approximate addition, absolute value subtraction, negation, minimum, and maximum, all realized using simple logical elements. Some operations like multiplication and approximate addition demand uncorrelated input bit-streams while others require correlated ones to obtain the correct results~\cite{CorrManipCirc}.

\subsection{Transcendental Functions 
}
\label{sc_transc_def}
Transcendental functions, such as $\sin(\cdot),\ \cos(\cdot),\ \exp(\cdot),\ \ln(\cdot)$,
and $\tanh(\cdot)$, are fundamental 
in many modern workloads, especially in signal and image processing, numerical kernel, and machine learning applications, where they are used for transforms, activation functions, and normalization~\cite{chen2023object, Wang2021}. Despite their importance, these operations are often treated as ``special functions'' in hardware: high-accuracy implementations typically rely on complex micro-architectures such as table-based methods, piecewise polynomial approximations, or iterative schemes. Such approaches incur substantial area and energy overhead and can become significant throughput bottlenecks when invoked repeatedly.

\begin{figure}[!t]
  \centering
  \includegraphics[width=0.7\textwidth]{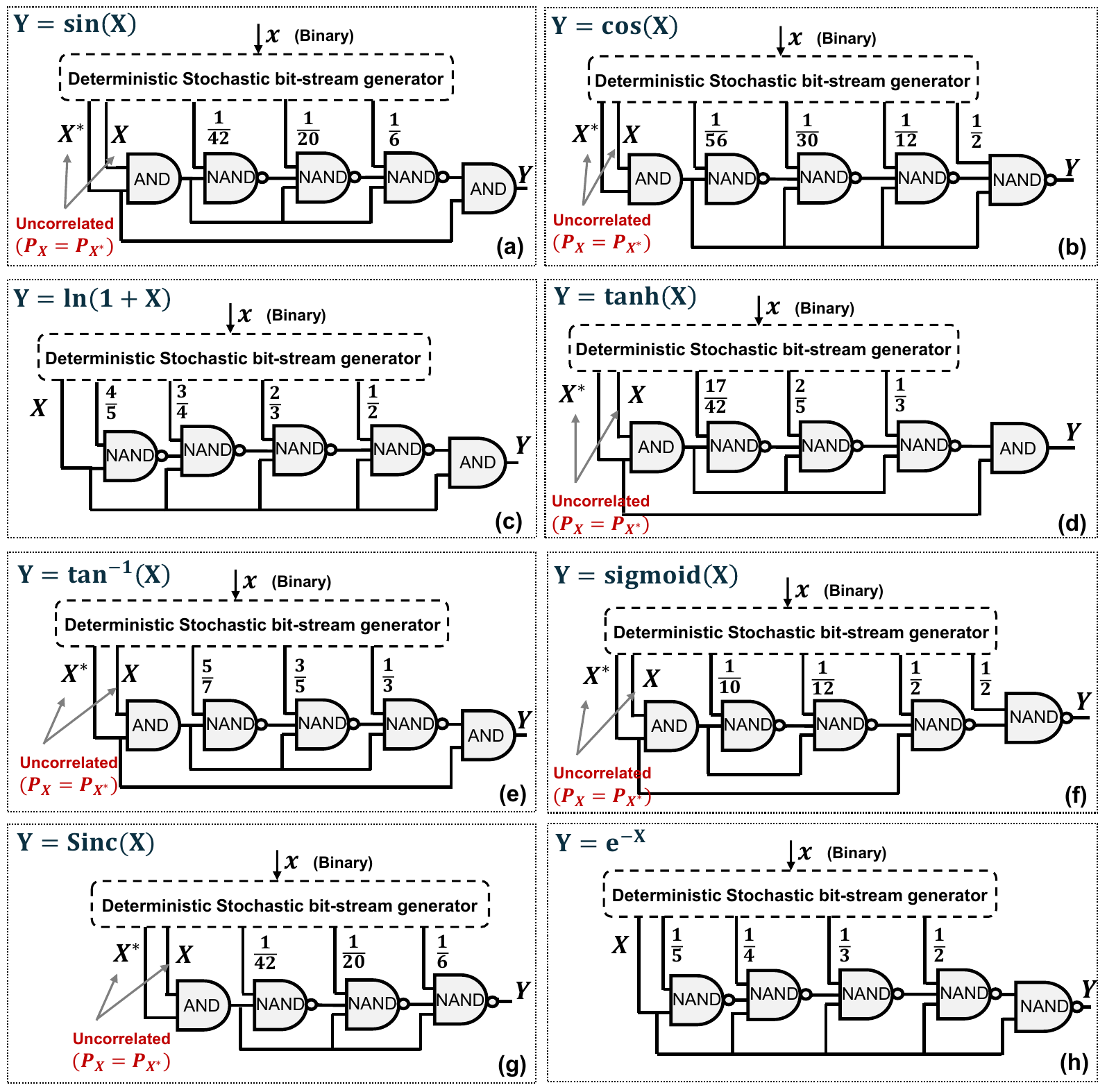}
  \caption{A general library of transcendental functions implementation using Stochastic logic.
  }
  \label{fig:sc_trans}
  \vspace{-.75em}
\end{figure}

This challenge is further amplified in IMC substrates. LIM fabrics favor regular, lightweight operations with limited control and short critical paths; as a result, conventional transcendental-function pipelines (with wide datapaths, multi-cycle iteration, and frequent intermediate state storage) can significantly deteriorate 
the benefits of reduced data movement. In contrast, SC offers a compelling alternative: it replaces wide arithmetic with simple logic operating on probabilistic bit-streams, while naturally supporting approximate computing in which accuracy is tuned by the bit-stream length. Most importantly for this work, SC aligns with the intrinsic strength of memory arrays---\textit{massive parallelism}. By generating and processing large bundles of stochastic bit-streams directly within the memory, the proposed approach enables concurrent evaluation of transcendental functions across many independent inputs, transforming 
traditionally expensive operators into throughput-oriented, array-parallel computations.

The Appendix Section
formalizes the implementation of transcendental functions using truncated Maclaurin-series expansions. These formulations are hardware-friendly, as they decompose each function into a structured composition of a small, reusable set of core SC arithmetic primitives. This formulation directly supports our objective: parallel, end-to-end in-memory execution of both arithmetic and transcendental operators using SC.
Prior SC-based implementations of transcendental functions typically rely on delay elements (\texttt{D-FF}s) at the input and/or intermediate stages of multi-level SC circuits to manage correlation~\cite{Trig-Parhi,9444648}. By employing deterministic bit-stream generation, our proposed approach completely eliminates the need for delay elements, resulting in purely combinational implementations, as depicted in Fig.~\ref{fig:sc_trans}. These SC circuits are naturally compatible with LIM structures, enabling scalable deployment across memory arrays and paving the way toward a massively parallel IMC engine.

\begin{figure}[t]
    \centering
    \includegraphics[width=0.5\textwidth]{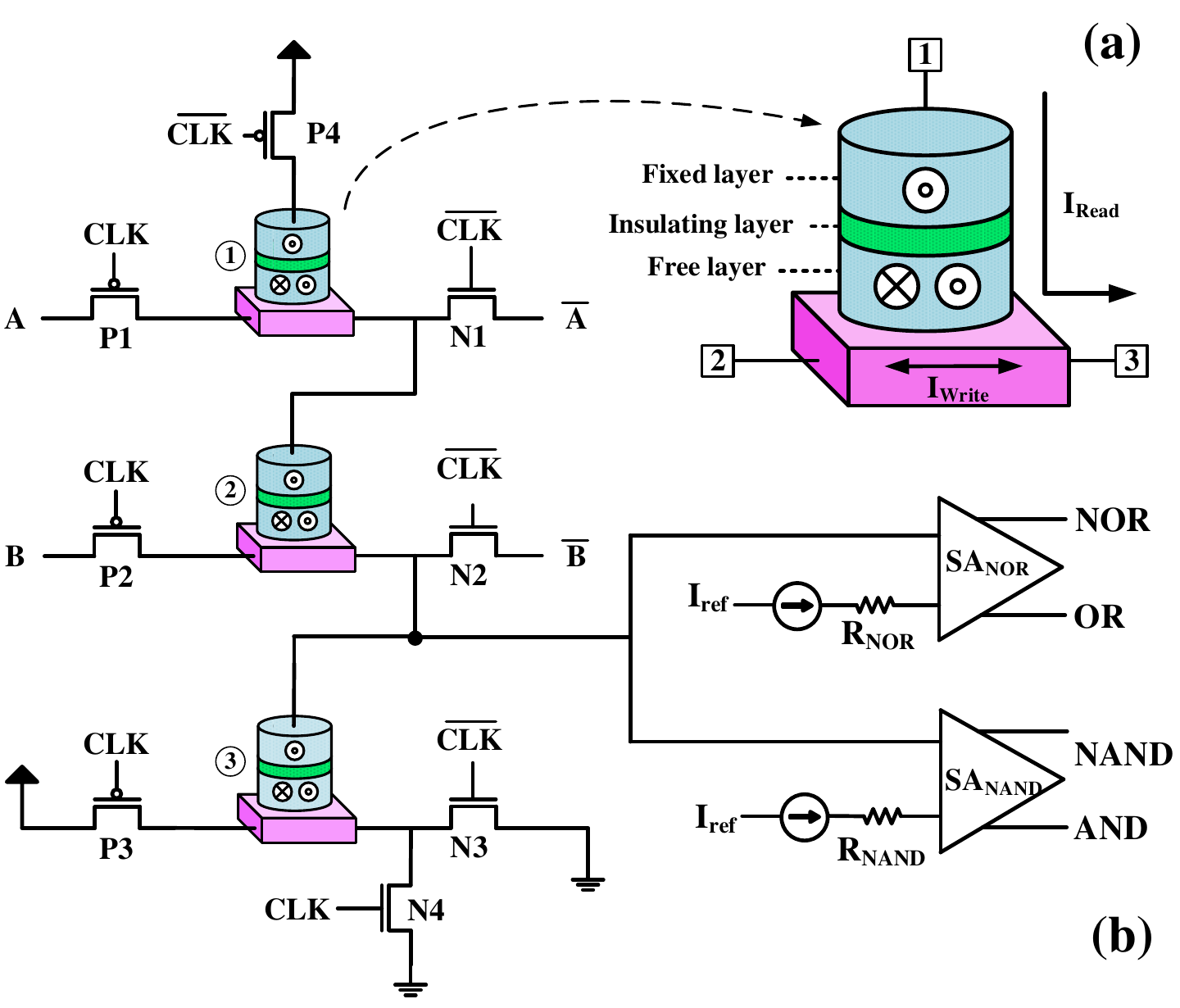}
    \caption{Structure of logic in-memory ~\cite{razi2022toward}, (a) MTJ device, (b) Circuit-level design.}
    \label{LIM}
    \vspace{-1.0em}
\end{figure}

\subsection{Magnetic Tunnel Junction (MTJ)}
\label{MTJ}

MTJs are non-volatile devices whose resistance is determined by the relative magnetic orientation of two ferromagnetic layers separated by a thin insulating barrier (Fig.~\ref{LIM}(a)), allowing binary information to be encoded using distinct resistance states. One magnetic layer maintains a fixed orientation, while the other can be switched by an applied current through a heavy metal, enabling reliable data storage and repeated read operations without state loss. When the magnetic orientations of the two layers are aligned in the parallel state, the MTJ exhibits a low resistance, whereas an antiparallel alignment results in a higher resistance. Beyond serving as memory elements, MTJs expose resistance-level information that can be directly sensed and compared, allowing them to participate in computation when combined with appropriate peripheral circuitry~\cite{Liu2012SpinHallTaSwitching}. This property enables LIM architectures in which simple logic functions are implemented within or near memory arrays. Moreover, MTJs are compatible with hybrid CMOS designs, allowing magnetic devices and conventional transistors to be integrated to form compact and efficient logic structures.

In this work, we adopt the LIM structure introduced in ~\cite{razi2022toward} as a computational primitive within the proposed in-memory stochastic architecture. The LIM (Fig.~\ref{LIM}(b)) is built around MTJ devices combined with lightweight CMOS peripheral circuits and operates in two clearly separated phases: (i) \textit{preparation} and (ii) \textit{evaluation}. During the \textit{preparation} phase, the LIM configures the magnetic states of MTJ1 and MTJ2 according to the input operands $A$ and $B$ read from memory, enabling controlled write currents to program each MTJ into the desired parallel or antiparallel state. This phase effectively encodes the input data into resistance states. In the subsequent \textit{evaluation} phase, the write paths are disabled and read currents are applied to the programmed MTJs, allowing their combined resistance to generate distinct voltage levels through voltage division. These voltage levels are then compared against reference levels using sense amplifiers (SAs) to produce the logic outputs. In our proposed architecture, this LIM structure is directly integrated into the memory subsystem and serves as the foundation for parallel stochastic operations.

\section{Proposed Architecture}
\label{prop}

\begin{figure*}[!t]
\centering
\includegraphics[width=0.85\linewidth]{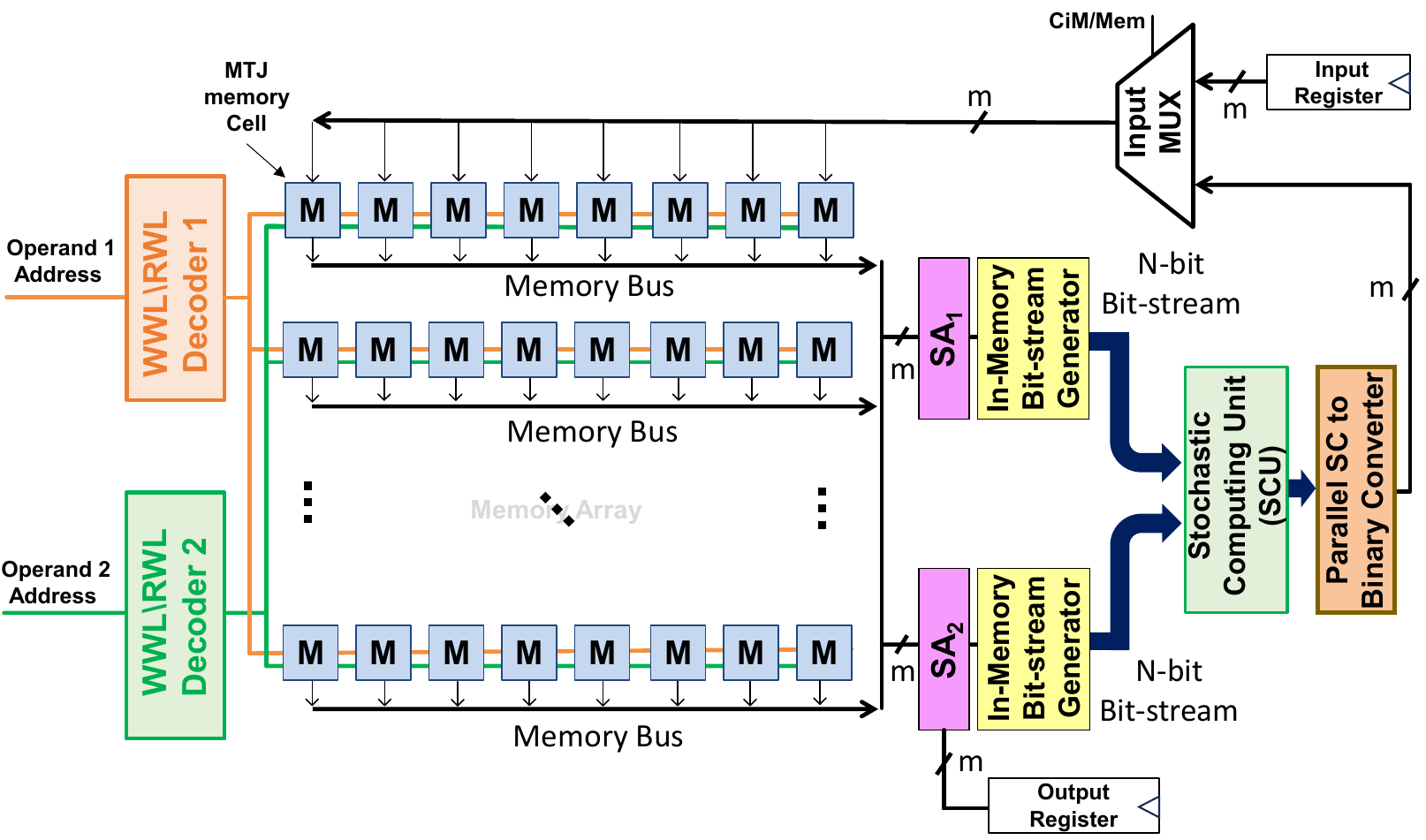}
\caption{General overview of proposed parallel in-memory stochastic architecture.}
\label{General}
\vspace{-1.5em}
\end{figure*}

In this section, we present our MTJ-based IMC architecture in which stochastic bit-stream generation, computation, and accumulation are unified into a single, end-to-end parallel pipeline via LIM.
Fig.~\ref{General} presents the overall organization and data flow of the proposed architecture. 
The operation of the memory array in the proposed architecture is controlled by a dedicated \texttt{CiM/Mem} control signal that dynamically switches the system between the memory storage mode and the computation mode. In memory mode, this control signal enables standard read and write operations, allowing data to be transferred between the magnetic memory array and external input or output registers, while all computation-related modules remain disabled. In this mode, an input multiplexer (\texttt{MUX}) selects data from the input register and forwards it to the memory array for conventional write operations. In computation mode, the same control signal activates the stochastic computing units (SCU) and associated conversion units, repurposing the memory array for in-memory stochastic computation.  
The input \texttt{MUX} then selects the output of the IMC pipeline, enabling computed results to be written back into the memory array without leaving the memory subsystem.

In the proposed structure, $m$-bit binary operands are first accessed from the magnetic memory array via dedicated wordline control logic and sensed by $m$-bit SAs. During computation mode, the sensed binary values are deterministically converted into $N = 2^m$ stochastic bit-streams by coordinating MTJ write and read operations across memory rows, allowing two bit-streams to be generated in parallel. These stochastic streams are then forwarded to $N$ parallel SCU implemented within or near the memory array, where arithmetic operations are executed concurrently across multiple data paths by the LIM structure. Finally, the stochastic results are converted back into binary form using parallel accumulation mechanisms and written back into the MTJ memory. By integrating storage, conversion, computation, and write-back within a unified and parallel in-memory framework, the proposed design minimizes data movement and exploits memory-level parallelism, providing an efficient substrate for data-intensive workloads. In the following subsections, we elaborate on the detailed operation of each module shown in Fig.~\ref{General}.


\begin{figure}[!t]
    \centering
    \includegraphics[width=0.7\textwidth]{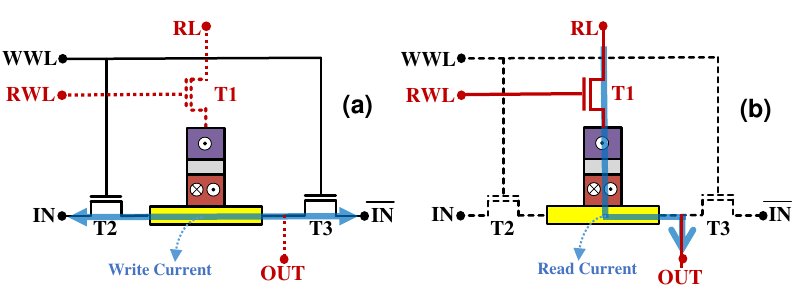}
    \vspace{-1em}
    \caption{MTJ-based Memory Cell, (a) Write, (b) Read Operations.}
    \label{MemCell}
\end{figure}

\subsection{Memory Structure}
\label{Memory}
The memory blocks labeled as \textit{M} in Fig.~\ref{General} represent the fundamental magnetic memory cells that form the MTJ-based memory array. Each memory cell consists of an MTJ device integrated with access transistors that enable independent read and write operations. The cell structure and its operating modes are illustrated in Fig.~\ref{MemCell}, which shows the peripheral connections and signal activation during write and read phases. The MTJ stores information in the form of magnetic orientation of its free layer, corresponding to distinct resistance states that represent binary data.

Write operations are performed by asserting the write wordline (WWL), which enables the write access path through the corresponding access transistors (shown in black). When WWL is activated, a write current is injected into the MTJ through transistors T2 and T3 to program its magnetic state, as illustrated in Fig.~\ref{MemCell}(a). The direction of the write current is determined by the input signal \textit{IN}, such that when $\textit{IN}=0$ the MTJ is configured into the parallel state, and when $\textit{IN}=1$ it is switched into the antiparallel state. This current-induced switching mechanism allows reliable programming of the cell while keeping the read path electrically isolated to avoid unintended sensing or disturbance. The write operation is coordinated by the address decoder at the row level, enabling multiple MTJs along the selected wordline to be programmed in parallel.

Read operations are performed independently of the write path by asserting the read wordline (RWL), which activates a separate read access transistor (shown in red). As shown in Fig.~\ref{MemCell}~(b), the read current flows from read line (RL) through transistor T1 and the MTJ along a dedicated sensing path. The resistance-dependent current developed across the MTJ is captured by an SA, which converts the analog signal into a digital binary value. The separation of read and write current paths enables repeated and parallel read operations without disturbing the stored magnetic state, a property that is essential for both conventional memory access and stochastic bit-stream generation.

At the array level, multiple memory cells share common wordlines and bitlines, enabling row-level and bank-level parallel access. When a row is selected, the resistance states of multiple MTJs are sensed simultaneously by parallel SAs, producing a vector of binary outputs in a single access. This organization naturally exposes memory-level parallelism and provides the foundation for our deterministic and parallel stochastic bit-stream generation.

\begin{figure}[!t]
    \centering
    \includegraphics[width=0.5\textwidth]{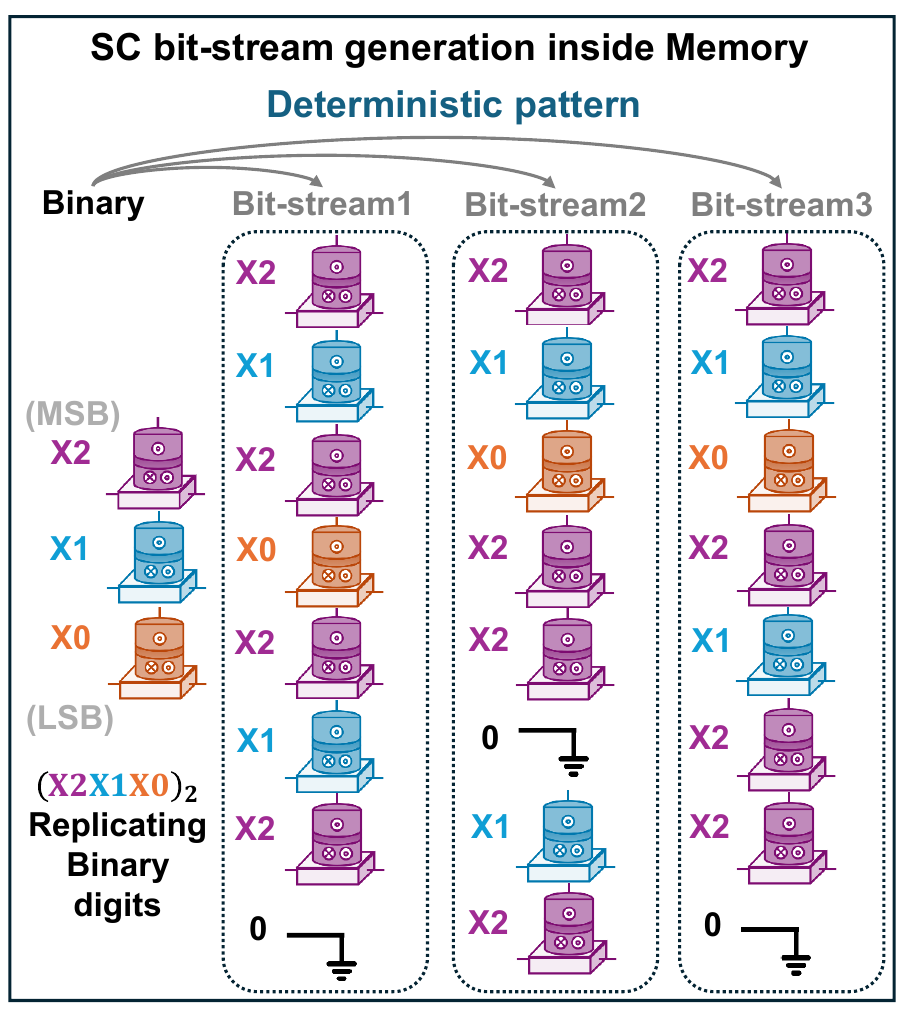}
    \vspace{-0.75em}
    \caption{An overview of in-memory deterministic SC bit-stream generation. The output bit-streams can be achieved through replicating binary digits using a deterministic pattern.
    }
    \label{fig:sc_imc_bs}
    \vspace{-1em}
\end{figure}

\subsection{In-Memory Bit-Stream Generation}
\label{RNG}
An essential stage in any SC system is the conversion of data from conventional binary to stochastic bit-streams. In the proposed architecture, bit-stream generation is performed directly inside the memory array using a deterministic bit-mapping mechanism. 
This mapping employs low-discrepancy, deterministic  patterns~\cite{Asadi_FSM_SBS_DATE21} that assign individual binary digits to specific positions within the bit-stream. 
Under this scheme, each binary bit $x_i$ is replicated $2^i$ times across the stream.
By selecting the same or distinct mapping patterns, the system can generate \textit{correlated} or \textit{independent} bit-streams, respectively. 
As an example, consider a 3-bit binary number $(x_2x_1x_0)_2$, which is mapped to a bit-stream of length $N=2^3=8$. In this case, $x_2$, $x_1$, and $x_0$ appear exactly $2^2$, $2^1$, and $2^0$ times, respectively. For instance, given the binary value  $(011)_2$, stochastic sequences such as ``$01010100$'' and ``$01100010$'' can be generated using the Bit-stream1 and Bit-stream2 patterns illustrated in Fig.~\ref{fig:sc_imc_bs}. 
This approach enables deterministic, fully parallel generation of bit-streams within the memory array.

\subsection{Parallel SC Unit}
\label{PSCU}
\begin{figure}[!t]
    \centering
    \includegraphics[width=1\textwidth]{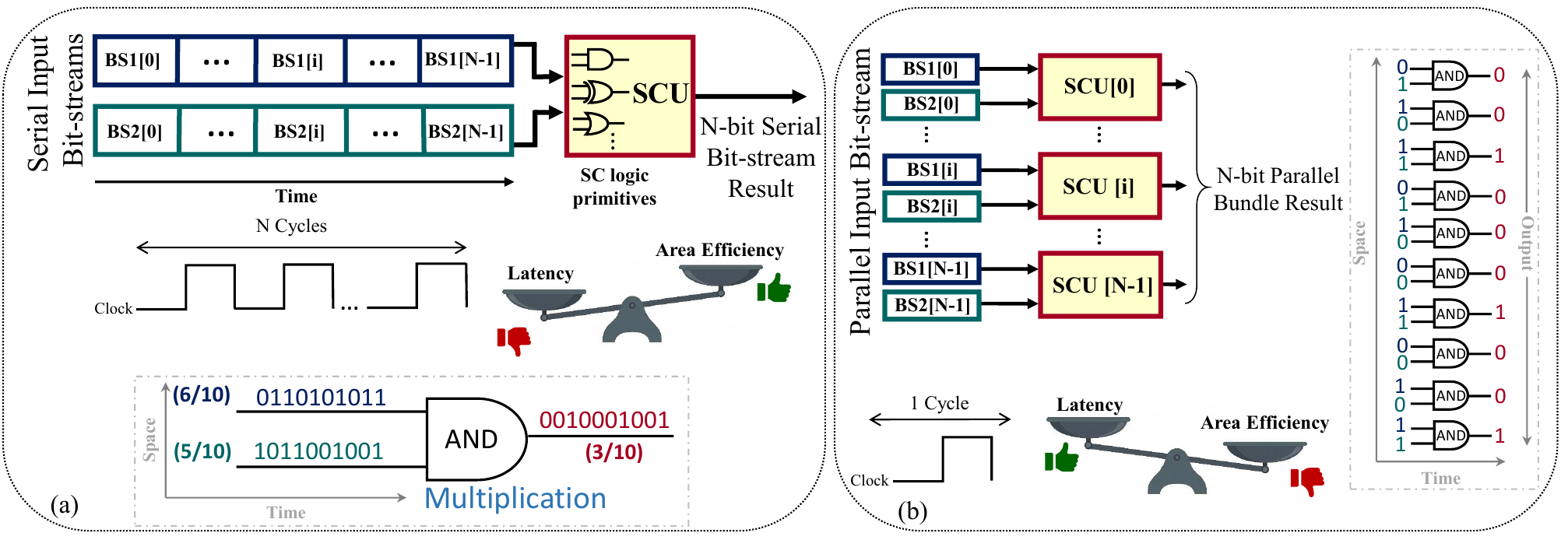}
    \caption{Stochastic Computing Unit (SCU) (a) Traditional Serial Approach, (b) Proposed Parallel Approach.}
    \label{FigSCU}
\end{figure}

The proposed architecture 
shifts SC from the traditional temporal, bit-serial processing to a spatial, massively parallel execution model. As illustrated in Fig.~\ref{FigSCU}(a), in traditional serial SC, a mathematical operation on a bit-stream of length $N$ requires $N$ clock cycles. In contrast, the proposed design leverages the massive internal bandwidth of the MTJ-based memory array to process the entire bit-stream concurrently. This execution model functionally mirrors the SIMD paradigm widely used in high-performance parallel computing, in which a single logic operation is applied simultaneously across a wide vector of data—here, the spatially distributed bits of the SC bit-streams (Fig.~\ref{FigSCU}(b)).
The proposed SC unit structure, utilizes the LIM to effectively transforms the memory array into a vast array of parallel processing elements that are shown in Figs.~\ref{fig:sc_op} and~\ref{fig:sc_trans}. Once binary operands are converted into SC bit-streams using the parallel generation method described in Section~\ref{RNG}, these streams no longer exist as sequences of events in time, but instead as a contiguous ``bundle'' of bits stored across memory rows or physically adjacent cells.

By activating the specific control signals for the LIM peripheral circuitry (e.g., enabling SAs for \texttt{NAND} or \texttt{NOR} logic), the architecture enforces a specific Boolean operation across the entire memory row. For instance, to perform 
multiplication (which maps to bit-wise logical \texttt{AND}), the system activates the \texttt{AND}-mode sensing across all column pairs holding the input bit-streams. This effectively executes $N$ bit-wise \texttt{AND} operations in parallel within a single memory cycle, regardless of the bit-stream length ($N$). This spatial unrolling allows the proposed architecture to execute essential SC arithmetic operations in constant time, i.e., $\mathcal{O}(1)$, independent of the bit-stream length, as all bits of the stochastic representation are processed concurrently within the memory array. Accordingly, this complexity characterization pertains to the core arithmetic execution within the memory array rather than the full end-to-end processing pipeline. This complexity characterization reflects the latency of the computation stage under a spatially parallel execution model. While the physical implementation of peripheral circuitry (e.g., sensing and interconnect structures) depends on array dimensions and technology parameters, the arithmetic operations themselves are executed in a constant number of cycles due to concurrent bit-wise processing within the memory array.

The architecture supports a comprehensive set of arithmetic operations realized through this spatial SIMD model. As shown in Fig.~\ref{fig:sc_op}, 
SC operations rely on standard logic elements. In our LIM array, these are executed by selecting the appropriate wordlines and sensing modes, generating the entire result vector instantly. Furthermore, operations requiring \texttt{XOR} gates (such as absolute value subtraction) are synthesized using a multi-step LIM sequence (e.g., combining \texttt{OR}, \texttt{NAND}, and \texttt{AND} steps) which are still broadcast in parallel across the array.
It is important to note that this parallel LIM approach applies to combinational stochastic logic. Operations that inherently require state-dependent feedback or temporal tracking, such as \textit{SC Division}~\cite{Moghadam_SC_Division_TNANO2024,Razi_IM_SC_Div_DAC25}, are not supported by this parallel combinatorial 
model and must be handled via sequential processing or external logic.

This SIMD-like capability is particularly effective 
for the \textit{transcendental} functions discussed in Section~\ref{sc_transc_def}, which are expressed using truncated Maclaurin series that reduce to cascaded chains of simple logic gates (\texttt{AND} and \texttt{NAND} gates).
Our deterministic, parallel bit-stream generation 
eliminates the need for temporal decorrelation elements (e.g.,  \texttt{D-FF}s) between successive logic stages. Consequently, these 
functions can be physically realized as combinational logic chains distributed across the memory peripherals. The memory controller orchestrates execution by issuing a sequence of LIM instructions (e.g., ``Step 1: Compute $x^2$ via \texttt{AND}'', ``Step 2: Scale result through a constant coefficient'', etc.) which propagate through the data path as illustrated in Fig.~\ref{fig:sc_trans}. Since each instruction operates simultaneously on the entire bit-vector, the complete transcendental evaluation is performed in only a few cycles (typically 2–3 cycles as reported in Table~\ref{tab:delay_power_comparison}), yielding orders-of-magnitude speedups over serial SC implementations and conventional iterative CORDIC-based approaches~\cite{CORDIC_TCASI_2009,singh2024energy}.

While the proposed spatial execution model enables constant-cycle evaluation of stochastic operations, this benefit is obtained by trading temporal serialization for spatial hardware parallelism. In conventional serial SC, an $N$-bit stochastic operation is typically executed using a single SC compute unit over $N$ cycles. In contrast, the proposed architecture spatially unrolls the stochastic bit-stream across the memory array and evaluates all $N$ bits concurrently in \underline{one} cycle. This parallelism is exploited at the bit level within each operation, rather than across multiple independent operands, and therefore primarily targets latency reduction of individual operations instead of increasing operand-level throughput. As a result, the computation stage replaces the temporal cost of $N$ serial evaluations with $N$ parallel SC evaluation paths and $m$-bit sensing interface, where $m$ denotes the memory data width, along with a modest overhead from control and peripheral circuitry. Therefore, relative to conventional serial SC, the proposed design reduces computation latency from $\mathcal{O}(N)$ to $\mathcal{O}(1)$ for the arithmetic stage, while increasing the required spatial compute and sensing resources approximately in proportion to the exposed parallelism.
This reflects a fundamental latency--hardware trade-off: higher parallelism reduces execution time, but requires wider sensing bandwidth, more concurrently active peripheral circuitry, and greater memory resource allocation for bit-stream placement. Conversely, when the available array width or peripheral budget is limited, the same architecture can operate with reduced parallelism, processing the bit-stream in multiple batches and recovering the corresponding latency scaling. Importantly, this trade-off is well aligned with memory-centric architectures, where large arrays already provide substantial row- and column-level concurrency. Thus, the proposed framework should be viewed as a scalable design space in which latency is reduced by exploiting available spatial resources, rather than as a claim of zero-cost parallelism.

\begin{figure}[!t]
    \centering
    \includegraphics[width=0.8\textwidth]{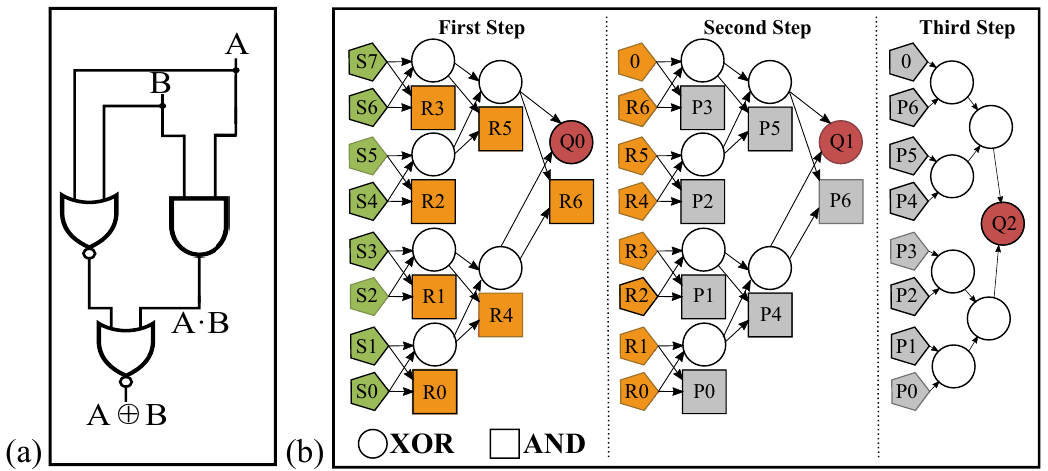}
    \caption{Parallel Bit-stream to binary conversion schematic used in~\cite{Alam_SC_Memristor_DT_2021}. (a) \texttt{XOR} operation using \texttt{NOR} gates and \texttt{AND} gate realized through LIM structure. (b) An example of conversion from 8-bit stream into 3-bit binary format $(Q_2Q_1Q_0)_2$ within three steps. Circles represent \texttt{XOR} and squares show \texttt{AND} operations.
    }
    \label{fig:sc_binary}
\end{figure}

\subsection{Parallel Bit-stream to Binary Converter}
\label{S2B}
The outputs produced by the LIM computation engine within 
the memory array can either be forwarded as inputs to subsequent computation stages 
or converted back to 
conventional binary format.
When a binary-format output is required, an additional bit-stream-to-binary conversion stage is performed. This conversion is achieved by accumulating the `1's in the bit-stream. 
To obtain the corresponding binary value from a given bit-stream, we adopt the method used in~\cite{Alam_SC_Memristor_DT_2021}, illustrated in Fig.~\ref{fig:sc_binary} for the representative case of 
generating a $log_2N=3$-bit binary output. 
This conversion process relies on a combination of \texttt{XOR} and \texttt{AND} logic operations. As shown in Fig.~\ref{fig:sc_binary}(a), the result of a logical \texttt{AND} operation between two input bits can be directly exploited to compute their logical \texttt{XOR}. This dependency enables each \texttt{AND–XOR} pair to be realized using two \texttt{NOR} gates and one \texttt{AND} gate. 
Assuming a latency of $1.5$ cycles for the \texttt{XOR} and $1$ cycle for the \texttt{AND} operation in our 
architecture, each step in Fig.~\ref{fig:sc_binary}(b) takes $4.5$ cycles. Therefore, the overall bit-stream to binary process takes $4.5\times log_2N$ cycles.

\section{Experimental Results}
\label{exp}

\subsection{Accuracy Analysis and Robustness}
To evaluate the accuracy and fault tolerance of the proposed architectures, we first conducted comprehensive software simulations in \texttt{MATLAB}. 
The Mean Absolute Error (MAE) of various arithmetic operations was measured using Monte-Carlo simulations, with results averaged over $100,000$ iterations. To 
assess robustness against soft errors, we injected varying levels of random noise into the inputs of both the SC and conventional binary implementations. To provide a fair comparison, for each case the binary precision is selected as $log_2N$-bit. Table~\ref{tab:sc_arith_noise_injection} summarizes the evaluation results. As expected, increasing the SC bit-stream length--analogous to increasing bit-precision in binary arithmetic--consistently improves 
accuracy, confirming the progressive precision property of SC. Moreover, the results demonstrate that, compared to their binary counterparts, SC-based operations exhibit significantly higher robustness as noise levels increase.

\begin{table*}[!t]
\centering
\caption{MAE (\%) Measurements of Fault-Tolerance: SC vs. Binary for Basic Arithmetic Operations}
\vspace{-1em}
\resizebox{\textwidth}{!}{
\begin{tabular}{l c *{14}{c}}
\toprule
\multirow{3}{*}{\begin{tabular}[c]{@{}c@{}} \textbf{Arithmetic}\\ \textbf{Operation} \end{tabular}} & \multirow{3}{*}{\begin{tabular}[c]{@{}c@{}}\textbf{Bit-stream}\\ \textbf{Length ($N$)} \end{tabular}} & \multicolumn{14}{c}{\textbf{Injected Noise (\%)}} \\
\cmidrule(lr){3-9} 
\cmidrule(lr){3-16}
& & \multicolumn{2}{c}{\textbf{0}} & \multicolumn{2}{c}{\textbf{1}} & \multicolumn{2}{c}{\textbf{2}} & \multicolumn{2}{c}{\textbf{5}} & \multicolumn{2}{c}{\textbf{10}} & \multicolumn{2}{c}{\textbf{20}} & \multicolumn{2}{c}{\textbf{30}} \\
\cmidrule(lr){3-4} \cmidrule(lr){5-6} \cmidrule(lr){7-8} \cmidrule(lr){9-10} \cmidrule(lr){11-12} \cmidrule(lr){13-14} \cmidrule(lr){15-16}
& & \textbf{SC} & \textbf{Binary} & \textbf{SC} & \textbf{Binary} & \textbf{SC} & \textbf{Binary} & \textbf{SC} & \textbf{Binary} & \textbf{SC} & \textbf{Binary} & \textbf{SC} & \textbf{Binary} & \textbf{SC} & \textbf{Binary} \\
\midrule

Multiplication & \multirow{5}{*}{16} & 3.89 & 0 & 4.16 & 3.72 & 4.47 & 4.37 & 5.42 & 6.22 & 7.13 & 9.08 & 10.5 & 13.9 & 13.7 & 17.7 \\
Scaled Addition &  & 3.77 & 0 & 4.28 & 3.82 & 4.74 & 4.48 & 5.99 & 6.37 & 7.81 & 9.21 & 10.9 & 13.9 & 13.7 & 17.7 \\
Abs. Subtraction &  & 2.09 & 0 & 3.29 & 4.39 & 3.34 & 4.67 & 7.30 & 8.19 & 11.4 & 13.1 & 18.0 & 19.9 & 22.8 & 23.9\\
Minimum &  & 3.19 & 0 & 3.51 & 3.86 & 3.87 & 4.62 & 5.14 & 6.75 & 7.36 & 10.1 & 11.7 & 15.8 & 15.4 &  20.3\\
Maximum &  & 3.06 & 0 & 3.66 & 3.93 & 4.23 & 4.70 & 5.81 & 6.92 & 8.19 & 10.2 & 12.3 & 15.8 & 15.9 & 20.4\\
\midrule

Multiplication & \multirow{5}{*}{64} & 1.06 & 0 & 1.41 & 1.66 & 1.79 & 2.49 & 2.96 & 4.85 & 4.86 & 8.28 & 8.46 & 13.7 & 11.9 & 17.9 \\
Scaled Addition &  & 0.95 & 0 & 1.46 & 1.66 & 1.90 & 2.50 & 3.03 & 4.86 & 4.71 & 8.22 & 7.86 & 13.5 & 11.0 & 17.5 \\
Abs. Subtraction &  & 0.52 & 0 & 1.66 & 2.08 & 2.65 & 3.50 & 5.44 & 7.41 & 9.62 & 12.7 & 16.5 & 19.8 & 21.5 & 24.0 \\
Minimum &  & 0.79 & 0 & 1.19 & 1.71 & 1.66 & 2.58 & 3.20 & 5.13 & 5.66 & 8.89 & 10.1 & 15.1 & 13.9 & 20.0 \\
Maximum &  & 0.78 & 0 & 1.38 & 1.70 & 1.95 & 2.60 & 3.53 & 5.15 & 5.95 & 8.97 & 10.3 & 15.2 & 14.1 & 20.1 \\

\bottomrule
\end{tabular}
}
\label{tab:sc_arith_noise_injection}
\end{table*}

Table~\ref{tab:Transc_noise_injection} extends our analysis to the implemented transcendental functions, including trigonometric functions ($\sin(x)$, $\cos(x)$, $\arctan(x)$), hyperbolic ($\tanh(x)$), and other 
nonlinear operators such as $Sinc(x)$, $sigmoid(x)$, exponential decay ($e^{-x}$), and natural logarithm ($\ln(1+x)$).
Under noise-free conditions (0\% injected noise), the results validate the accuracy of the truncated Maclaurin-series implementations. As the SC bit-stream length increases from $N=16$ to $N=64$, a consistent reduction in MAE is observed across all functions. 
For example, at $N=64$, complex functions such as $\cos(x)$, $Sinc(x)$, and $sigmoid(x)$ achieve high fidelity, with MAE values dropping to approximately $0.8\%$, $0.9\%$, and $1.1\%$, respectively. These results demonstrate that the proposed purely combinational LIM structures (Fig.~\ref{fig:sc_trans}) can accurately approximate complex transcendental functions without relying on sequential state elements.

\begin{table*}[t]
\centering
\caption{MAE (\%) Measurements of Fault-Tolerance: SC vs. Binary for Transcendental Functions}
\vspace{-1em}
\renewcommand{\arraystretch}{1.1} 
\resizebox{\textwidth}{!}{
\begin{tabular}{l c *{14}{c}}
\toprule
\multirow{3}{*}{\textbf{Function}} & \multirow{3}{*}{\shortstack[c]{\textbf{Bit-stream} \\ \textbf{Length ($N$)}}} & \multicolumn{14}{c}{\textbf{Injected Noise (\%)}} \\
\cmidrule(lr){3-16}
& & \multicolumn{2}{c}{\textbf{0}} & \multicolumn{2}{c}{\textbf{1}} & \multicolumn{2}{c}{\textbf{2}} & \multicolumn{2}{c}{\textbf{5}} & \multicolumn{2}{c}{\textbf{10}} & \multicolumn{2}{c}{\textbf{20}} & \multicolumn{2}{c}{\textbf{30}} \\
\cmidrule(lr){3-4} \cmidrule(lr){5-6} \cmidrule(lr){7-8} \cmidrule(lr){9-10} \cmidrule(lr){11-12} \cmidrule(lr){13-14} \cmidrule(lr){15-16}
& & \textbf{SC} & \textbf{Binary} & \textbf{SC} & \textbf{Binary} & \textbf{SC} & \textbf{Binary} & \textbf{SC} & \textbf{Binary} & \textbf{SC} & \textbf{Binary} & \textbf{SC} & \textbf{Binary} & \textbf{SC} & \textbf{Binary} \\
\midrule

$\sin(x)$ & \multirow{8}{*}{16} & 5.33 & 0 & 5.33 & 5.89 & 5.42 & 6.60 & 5.88 & 8.64 & 7.40 & 11.7 & 11.0 & 17.0 & 14.9 & 21.7 \\
$\cos(x)$ & & 2.36 & 0 & 2.52 & 3.27 & 2.71 & 3.63 & 3.25 & 4.82 & 4.30 & 6.50 & 6.71 & 9.56 & 9.14 & 12.0 \\
$\tanh(x)$ & & 4.95 & 0 & 5.28 & 5.26 & 5.55 & 5.87 & 6.41 & 7.68 & 7.96 & 10.3 & 10.9 & 15.3 & 14.2 & 19.3 \\
$\arctan(x)$ & & 3.88 & 0 & 4.38 & 5.16 & 4.88 & 5.77 & 6.25 & 7.58 & 8.56 & 10.4 & 13.3 & 14.9 & 17.7 & 19.1 \\
$Sinc(x)$ & & 1.60 & 0 & 1.71 & 1.13 & 1.82 & 1.25 & 2.13 & 1.63 & 2.61 & 2.22 & 3.45 & 3.25 & 4.22 & 4.13 \\
$sigmoid(x)$ & & 2.89 & 0 & 2.98 & 1.63 & 3.09 & 1.83 & 3.35 & 2.36 & 3.79 & 3.23 & 4.78 & 4.70 & 5.59 & 5.90 \\
$\exp(-x)$ & & 2.19 & 0 & 2.58 & 4.65 & 2.98 & 5.16 & 4.09 & 6.58 & 5.78 & 8.92 & 8.93 & 12.9 & 12.0 & 16.4 \\
$\ln(1+x)$ & & 3.44 & 0 & 3.68 & 5.50 & 3.94 & 6.12 & 4.70 & 7.76 & 6.11 & 10.5 & 9.21 & 15.2 & 12.7 & 19.2 \\
\midrule


$\sin(x)$ & \multirow{8}{*}{64} & 2.40 & 0 & 2.36 & 2.11 & 2.47 & 2.88 & 3.24 & 5.18 & 5.04 & 8.74 & 9.12 & 15.0 & 13.3 & 20.3 \\
$\cos(x)$ & & 0.84 & 0 & 1.00 & 1.17 & 1.18 & 1.59 & 1.80 & 2.83 & 2.92 & 4.82 & 5.21 & 8.18 & 7.54 & 11.1 \\
$\tanh(x)$ & & 2.19 & 0 & 2.40 & 1.87 & 2.64 & 2.61 & 3.54 & 4.59 & 5.18 & 7.71 & 8.68 & 13.4 & 12.3 & 17.9 \\
$\arctan(x)$ & & 2.26 & 0 & 2.47 & 1.86 & 2.80 & 2.48 & 4.03 & 4.55 & 6.49 & 7.67 & 11.6 & 13.2 & 16.4 & 17.8 \\
$Sinc(x)$ & & 0.87 & 0 & 0.89 & 0.40 & 0.92 & 0.55 & 0.99 & 0.98 & 1.12 & 1.63 & 1.43 & 2.82 & 1.79 & 3.77 \\
$sigmoid(x)$ & & 1.08 & 0 & 1.14 & 0.58 & 1.20 & 0.80 & 1.42 & 1.42 & 1.83 & 2.40 & 2.83 & 4.08 & 3.94 & 5.50 \\
$\exp(-x)$ & & 1.38 & 0 & 1.48 & 1.67 & 1.63 & 2.24 & 2.25 & 3.94 & 3.65 & 6.64 & 6.82 & 11.4 & 10.1 & 15.3 \\
$\ln(1+x)$ & & 1.79 & 0 & 1.86 & 1.94 & 1.98 & 2.65 & 2.62 & 4.65 & 4.10 & 7.69 & 7.61 & 13.2 & 11.2 & 17.7 \\

\bottomrule
\end{tabular}
}
\label{tab:Transc_noise_injection}
\end{table*}

Table~\ref{tab:Transc_noise_injection} further highlights the superior fault tolerance of the SC implementation under increasing soft-error rates. As noise injection levels rise from 5\% to 30\%, the conventional binary implementations experience rapid and often catastrophic accuracy degradation. For instance, at 30\% noise level, the binary error for $\sin(x)$ increases to $21.7\%$, while for $\tanh(x)$ reaches $19.3\%$. In 
contrast, the proposed parallel SC architecture exhibits a 
more graceful degradation in performance. Under the same 30\% noise condition, the SC implementation maintains 
lower error levels, with $\sin(x)$ and $\tanh(x)$ errors of $14.9\%$ and 
$14.2\%$, respectively, even with a significantly shorter bit-stream length of $N=16$.
This resilience is particularly pronounced for activation functions commonly used in neural networks. For the $sigmoid(x)$ function at 20\% noise injection, the binary error ($4.08\%$) is 
higher than the SC error ($2.83\%$) at $N=64$. This gap confirms that the distributed representation of SC bit-streams—where no single bit carries significant weight—provides intrinsic immunity to soft errors. Consequently, the proposed architecture not only enables massive parallelism but also offers inherent reliability for noise-prone environments and approximate computing applications where robust operation is critical.

\begin{table*}[!t]
\centering
\vspace{1em}
\caption{Circuit-Level Measurements of Serial vs. Parallel In-Memory SC Implementations
}
\vspace{-0.75em}
\label{tab:delay_power_comparison}
\renewcommand{\arraystretch}{1.1}

\resizebox{\textwidth}{!}{%
\begin{tabular}{l l c c c c c c c c c c c c}
\toprule
\textbf{Metric} & \begin{tabular}[l]{@{}l@{}} \textbf{Operation}\\ \textbf{Mode} \end{tabular}
& \begin{tabular}[c]{@{}c@{}} \textbf{Multiplication}\\ \& \textbf{Minimum} \end{tabular} & \textbf{Maximum} & \begin{tabular}[c]{@{}c@{}} \textbf{Scaled}\\ \textbf{Addition} \end{tabular} & \begin{tabular}[c]{@{}c@{}} \textbf{Abs.} \\ \textbf{Subtraction} \end{tabular}
& $\mathbf{sin(x)}$ & $\mathbf{cos(x)}$ & $\mathbf{tanh(x)}$ & $\mathbf{arctan(x)}$
& $\mathbf{Sinc(x)}$ & $\mathbf{sigmoid(x)}$
& $\mathbf{exp(-x)}$ & $\mathbf{ln(1+x)}$ \\
\midrule

\multirow[c]{2}{*}{\begin{tabular}[c]{@{}c@{}} \textbf{Delay}\\ (Cycles) \end{tabular}}
& \textbf{Parallel}  & 1  & 1  & 2  & 2  & 3  & 3  & 3  & 3  & 2.5 & 3  & 3  & 3  \\
& \textbf{Serial}  & 64 & 64 & 66 & 66 & 67 & 67 & 67 & 67 & 67  & 67 & 67 & 67 \\
\midrule

\multirow[c]{2}{*}{\begin{tabular}[c]{@{}c@{}} \textbf{Power}\\ ($mW$) \end{tabular}}
& \textbf{Parallel}  & 3.1  & 3.1  & 9.3   & 9.3   & 15.4 & 15.4 & 15.4 & 15.4 & 12.3 & 15.2 & 15.3 & 15.3 \\
& \textbf{Serial} & 0.05 & 0.05 & 0.15 & 0.15 & 0.87 & 0.62 & 0.87 & 0.87 & 0.44 & 0.62 & 0.74 & 0.74 \\
\midrule

\multirow[c]{2}{*}{\begin{tabular}[c]{@{}c@{}} \textbf{Power $\times$ Delay}\\  ($mW \times Cycles$) \end{tabular}}
& \textbf{Parallel}  & 3.1 & 3.1 & 18.7 & 18.7 & 46.3 & 46.1 & 46.2 & 46.2 & 30.8 & 45.6 & 46.0 & 45.9 \\
& \textbf{Serial}  & 3.2 & 3.2 & 9.9  & 9.9  & 58.3 & 41.6 & 58.3 & 58.3 & 29.5 & 41.5 & 49.6 & 49.6 \\
\bottomrule
\end{tabular}
}
\end{table*}

\subsection{Circuit-Level Analysis}
We next evaluate the functional behavior and performance characteristics of the proposed parallel in-memory 
architecture through detailed HSPICE circuit-level simulations. A representative set of stochastic operations is examined within the MTJ-based memory system, including basic arithmetic functions (multiplication, addition, and subtraction) as well as a range of transcendental functions. 
Both serial and parallel 
execution models were evaluated to quantify the performance impact of architectural parallelism. All simulations were performed using a hybrid CMOS--magnetic technology framework ~\cite{razi2022toward}, in which the peripheral circuitry, including access devices, SAs, and control logic, is implemented using predictive PTM FinFET models, while the memory and LIM functionality are realized using calibrated in-plane magnetic tunnel junction (i-MTJ) device models~\cite{kang2016spintronic}. A bit-stream length of 64 was used across all evaluated functions to ensure consistent computational accuracy. In the parallel configuration, a parallelism level of 64 was employed to fully exploit concurrent bit-stream processing within the memory array. All results are obtained at a supply voltage of \(V_{DD}=0.9\,\text{V}\), enabling fair and consistent comparison across different execution modes and function types.

Table~\ref{tab:delay_power_comparison} quantitatively compares the latency, power consumption, and power--delay product (PDP) of parallel and serial execution across both arithmetic and transcendental functions implemented in the proposed 
framework. The results clearly demonstrate the latency advantage of parallel execution. While serial implementations require on the order of 64--67 cycles to process a complete stochastic bit-stream, the proposed parallel architecture completes basic arithmetic operations such as multiplication and addition in only 
1--2 cycles, and more complex transcendental functions (e.g., $\sin(x)$, $\tanh(x)$, $\arctan(x)$) within three cycles. 
As discussed in Section~\ref{Related_Works} and summarized in Table~\ref{tab:latency_comparison}, prior IMC approaches typically incur \underline{thousands of cycles} for arithmetic and nonlinear operations due to bit-serial execution, iterative accumulation, or stateful memory operations, highlighting the substantial latency advantage of the proposed parallel 
in-memory design.

From a power perspective, parallel execution incurs higher instantaneous power consumption than its serial counterpart due to increased concurrent activity. However, when delay and power are jointly considered through the PDP metric, the parallel architecture consistently outperforms or closely matches serial execution across both arithmetic and transcendental functions. In particular, the significant reduction in execution cycles effectively compensates for the higher parallel power, yielding favorable delay--power trade-offs for complex nonlinear functions where serial SC suffers from large temporal overheads. 
As discussed in Section~\ref{sc_transc_def} conventional serial SC implementations rely on delay elements (e.g., D-FFs) for correlation management and temporal alignment, which introduce extra switching activity and sequential overhead. By largely eliminating these delay elements through combinational in-memory execution, the proposed architecture reduces per-bit switching overhead, further contributing to improved PDP.
Overall, the results in Table~\ref{tab:delay_power_comparison} confirm that exploiting parallelism inside memory is essential for achieving scalable SC with low latency and manageable power, especially beyond basic arithmetic operations.

\begin{table*}[t]
\centering
\caption{Comparison with State-of-the-Art In-Memory Computing Architectures}
\vspace{-0.75em}
\renewcommand{\arraystretch}{1.5}
\label{tab:latency_comparison}
\resizebox{\textwidth}{!}{%
\begin{tabular}{l c c c c c c c}
\toprule
\textbf{Method} & \textbf{Technology} & \textbf{Logic Paradigm} & \textbf{Arithmetic} & \textbf{Transcendental} & \textbf{Exec. Model} & \textbf{Latency} & \textbf{Power / Energy} \\ 
\midrule

\textcolor{Navy}{\textbf{TCAS-I'22}~\cite{lakshmi2022inner}} & ReRAM & Distributed Arithmetic & Inner Product & No & Seq. Accumulation & $\sim$1,458 cycles (64b) & $\sim$0.3 pJ/op \\ 

\textcolor{Navy}{\textbf{JSSC'19}~\cite{wang201928}} & SRAM & Bit-Serial CRAM & Add, Sub, Mul & No & Bit-Serial & $\sim$4,414 cycles (64b) & $\sim$19 pJ/bit \\ 

\textcolor{Navy}{\textbf{D\&T'21}~\cite{Alam_SC_Memristor_DT_2021}} & Memristor & Exact SC (MAGIC) & Exact Mul & No & Stateful Logic & 3 cycles (Full precision) & 90 pJ (6-bit precision) \\

\textcolor{Navy}{\textbf{ISCAS'18}~\cite{haj2018efficient}} & Memristor & MAGIC & Mul (Fixed) & No & Seq. Stateful & $\sim$52,358 cycles (64b) & $\sim$0.1 pJ/bit \\

\textcolor{Navy}{\textbf{ISVLSI'24}~\cite{singh2024energy}} & DRAM & CORDIC & MAC & Yes (Iterative) & Iterative Shifts & $\sim$375 cycles (Avg) & $\sim$1.2 nJ/op \\

\textcolor{Navy}{\textbf{ISLPED'25}~\cite{kim2025memraptor}} & MRAM & LUT / ROM & MVM & Yes (Table) & Memory Lookup & $\sim$1-5 cycles & $\sim$0.5 pJ/bit \\

\textcolor{Navy}{\textbf{Micromachines'21}~\cite{xu2021high}} & CIM & Analog/Mixed & MAC & Sigmoid Only & Circuit-Specific & $\sim$1 cycle (Analog) & $\sim$3 $\mu$W \\ 

\textcolor{Navy}{\textbf{ISPASS'23}~\cite{TransPimLib}} & DRAM & Software / ISA & All (Soft) & Yes (Library) & Core Instruction & $>$ 1,000 cycles & High (Core Active) \\ 

\textcolor{Navy}{\textbf{JXCDC'23}~\cite{zink2023stochastic}} & MRAM & TRNG SC (CRAM) & MAC / Add & No & In-Memory SC & $O(N)$ cycles & High Efficiency \\

\rowcolor[HTML]{EFEFEF}
\textbf{This Work} & \textbf{MRAM} & \textbf{Deterministic Parallel SC} & \textbf{All Basic} & \textbf{Universal} & \textbf{Spatial SIMD} & \textbf{1 - 3 cycles} & \textbf{3.1 -- 15.4 mW} \\ 
\bottomrule
\end{tabular}%
}
\vspace{-1em}
\end{table*}

\subsection{Comparison with Related Works}
\label{Related_Works}
Several prior works have explored arithmetic execution within memory using diverse architectural and technological approaches. The authors in \cite{lakshmi2022inner} employ distributed arithmetic to realize in-memory inner-product computation by storing precomputed coefficient combinations in a ReRAM array and performing accumulation through majority-logic-based adder--shifter circuits, eliminating explicit multipliers but requiring repeated read and write operations that enforce sequential accumulation. A different direction is taken in~\cite{wang201928}, which proposes a compute-SRAM (CRAM) architecture in scaled CMOS that executes arithmetic in a bit-serial manner using near-memory logic, support for multiple arithmetic functions, albeit with fundamentally serial bit-level execution, and frequent intermediate write-backs. 
Focusing on non-volatile memories, \cite{alam2020exact} introduces an exact in-memory multiplication scheme based on deterministic SC implemented via memristor-aided logic (MAGIC), where binary operands are transformed into deterministic stochastic bit-streams and processed using stateful logic operations, guaranteeing numerical exactness at the expense of long bit-stream processing. Building on MAGIC-based computation,~\cite{haj2018efficient} proposes a fixed-point in-memory multiplication technique that generates and accumulates partial products inside memristive arrays, leveraging row-level parallelism but still relying on sequential stateful operations that result in quadratic scaling with operand precision. More broadly,~\cite{imani2017ultra} presents a general IMC framework that combines in-memory and near-memory logic to support multiple arithmetic primitives; however, its computation model remains inherently sequential and does not fundamentally improve the latency scaling of multi-bit arithmetic operations.
More recently, Zink et al.~\cite{zink2023stochastic} proposed SC-CRAM, an architecture that leverages the thermal noise of MTJs to perform True Random Number Generation (TRNG) for SC within the memory array. While highly energy-efficient for basic MAC operations, SC-CRAM processes bit-streams temporally, resulting in execution times proportional to the stochastic sequence length ($O(N)$ cycles), and it lacks native structural support for evaluating complex transcendental functions.

Table~\ref{tab:latency_comparison} illustrates an overall comparison with other related IMC works implementing arithmetic and transcendental functions.
The distributed-arithmetic-based inner-product engine in \cite{lakshmi2022inner} exhibits cycle counts that grow proportionally with both operand width and accumulation length, reaching on the order of thousands of cycles at 64-bit precision. Bit-serial compute-SRAM designs such as \cite{wang201928} achieve relatively low latency for addition and subtraction at small bit widths; however, multiplication latency increases quadratically with precision, exceeding several thousand cycles for 64-bit operands. 
Finally, the general IMC framework in \cite{imani2017ultra} supports multiple arithmetic primitives but still demonstrates linear or quadratic growth in cycle count, leading to non-negligible latency at higher precisions.

Furthermore, recent work has explored support for transcendental functions in IMC systems using diverse architectural approaches. The authors in \cite{singh2024energy} integrate transcendental computation into DRAM-based IMC using iterative CORDIC algorithms to realize functions such as sine, exponential, and logarithm with tunable precision. \cite{xu2021high} demonstrates circuit-level realization of specific transcendental functions, such as sigmoid, using in-memory or near-memory computing primitives, highlighting the feasibility of implementing nonlinear functions directly within memory-centric architectures. Complementary studies quantify the performance cost of transcendental computation in PIM systems: \cite{TransPimLib} reports that transcendental function implementations on programmable PIM cores typically require thousands of execution cycles per input element, while \cite{singh2024energy} and \cite{kim2025memraptor} evaluate latency indirectly through throughput and normalized execution time. Overall, these works demonstrate the feasibility of transcendental functions in PIM systems but indicate that latency remains significant due to iterative computation, lookup overhead, or software execution, motivating more direct in-memory transcendental implementations.

Overall, despite their individual advantages, 
prior approaches suffer from long execution latency and high cycle counts 
when operating at moderate to high precision. 
To address these limitations, we design an end-to-end in-memory SC architecture that exploits parallelism across the memory array, enabling significantly reduced effective latency while supporting a broader class of arithmetic and transcendental operations.

\subsection{PIM-PIM: Processing-in-Memory Parallel Image Processing}
\label{Benchmark}
In this section, we benchmark the proposed MTJ-based stochastic function blocks in an image processing workload, referred to as \textit{PIM-PIM} (\textbf{P}rocessing-\textbf{I}n-\textbf{M}emory \textbf{P}arallel \textbf{IM}age Processing). The key motivation is to overcome a long-standing limitation of SC: 
computation latency~\cite{8782977}. Conventional serial SC evaluates a function over a bit-stream of length $N$ by producing and processing one bit per cycle, resulting in $O(N)$ latency per pixel and severely limiting throughput in practical workloads. 
In contrast, PIM-PIM bridges the proposed device-level MTJ primitives with an architecture-level, in-memory image-processing pipeline, enabling a systematic evaluation of 
their end-to-end impact on both output quality and hardware efficiency for representative pre-processing kernels used in learning systems.

We choose tone mapping~\cite{10.1145/566654.566574} as the target application for evaluation because it is (i) embarrassingly parallel, with each pixel processed independently, and (ii) well suited to exposing 
the behavior of the proposed non-linear function blocks. In particular, tone mapping is naturally expressed using an  S-shaped transfer characteristic based on either a $sigmoid$ or tanh nonlinearity, making it a suitable benchmark for our 
in-memory designs. Given a normalized input intensity $x \in [0,1]$, the output intensity $y$ is computed as $y = f(x)$ where $f(\cdot)\in\{sigmoid(\cdot), \tanh(\cdot)\}$. In our experiments, the nonlinear function $f(\cdot)$ is applied independently to each pixel to produce the output image.

In tone mapping, the input intensity is first centered and scaled around a tunable parameter $\beta$ and then clipped to the valid dynamic range according to $x' = \mathrm{clip}\ \left(\beta + \alpha(x-\beta),\,0,\,1\right)$, where $\alpha$ adjusts the contrast gain (larger $\alpha$ yields a steeper response around $\beta$) and $\beta$ defines the transition point (typically $\beta\ =\ 0.5$ for mid-gray). We then apply one of the two nonlinear mappings:
\begin{align}
y_{sigmoid} &= \sigma\!\left(k(x' - c)\right), \label{eq:tone_sigmoid}\\
y_{\tanh}  &= \tfrac{1}{2}\Big(\tanh\!\left(k(x' - c)\right) + 1\Big), \label{eq:tone_tanh}
\end{align}
where $k$ controls the slope (i.e., the strength of the S-curve) and $c$ determines the center of the curve. Intuitively, these S-shaped transfer functions amplify differences in mid-tone intensities while compressing extreme shadows and highlights, enabling contrast enhancement without hard saturation.

PIM-PIM exploits two levels of parallelism enabled by IMC. First, it employs bit-parallel, spatially unrolled stochastic bit-streams; rather than generating and consuming $N$ bits sequentially, an $N$-way parallel representation is realized across memory cells, allowing each in-memory logic operation to operate on all $N$ bits simultaneously. Second, PIM-PIM leverages pixel-level parallelism across the memory array. Given $C_{\text{total}}$ usable memory columns, the architecture can process $P = \left\lfloor \frac{C_{\text{total}}}{N} \right\rfloor$ pixels concurrently per batch. 

\begin{figure*}[t]
    \centering
    \includegraphics[width=\textwidth]{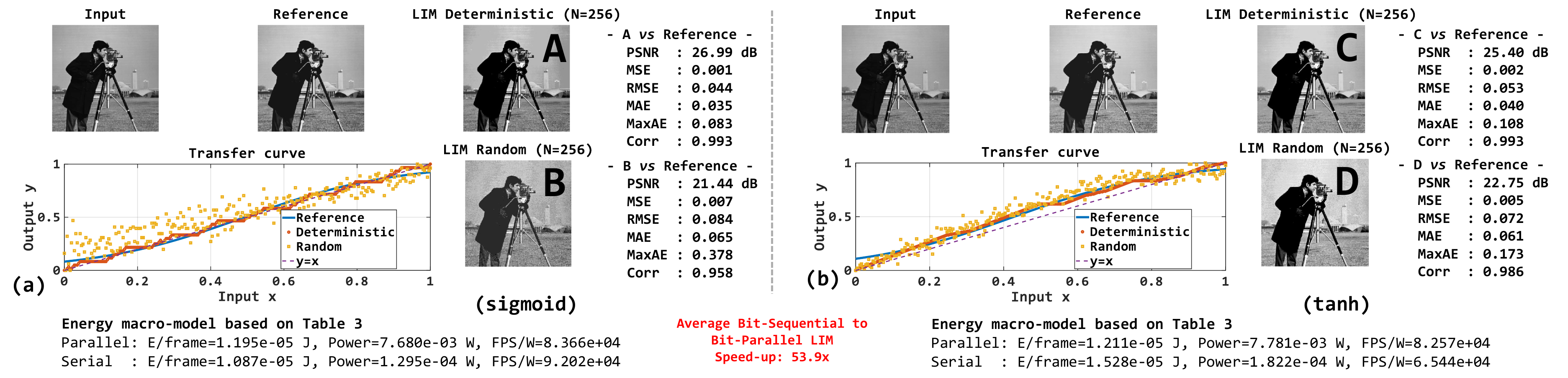}
    \vspace{-1.8em}
    \caption{PIM-PIM tone-mapping benchmark using the proposed in-memory (a) $sigmoid$ block, and (b) tanh block. Input image and reference output obtained from a smooth best-fit binary-processing $sigmoid$ curve. PIM-PIM outputs using deterministic and random bitstreams, annotated with image-level quality metrics versus the reference. Transfer characteristic ($x\!\rightarrow\!y$) showing the fitted reference curve, deterministic and random models, and the identity line ($y$=$x$). Deterministic streams closely track the reference tone curve and achieve higher quality. 
    {\footnotesize{\textit{PSNR: Peak Signal-to-Noise Ratio, MSE: Mean Squared Error, RMSE: Root Mean Squared Error, MAE: Mean Absolute Error, MaxAE: Maximum Absolute Error, Corr: Pearson Correlation Coefficient, E/frame: Energy per frame, FPS: Frames per second. Input image size: 256-by-256.}}}}
    \label{pimpim1}
\end{figure*}

Fig.~\ref{pimpim1} presents the PIM-PIM results for tone mapping using the in-memory (a) $sigmoid$ and (b) tanh blocks at $N=256$ with parameters $(\alpha,\beta)=(1.2,0.5)$. The deterministic implementation produces an enhanced image that closely 
matches the reference, 
while the random implementation introduces noticeable grain and intensity distortion. Quantitatively, for the $sigmoid$ function in Fig.~\ref{pimpim1}~(a), deterministic streams reach 26.99~dB PSNR, whereas random streams drop to 21.44~dB PSNR. Similarly, for $\tanh$ in Fig.~\ref{pimpim1}~(b), deterministic streams achieve 25.40~dB PSNR, while the random case yields 22.75~dB PSNR. The transfer-curve subplots explain this gap: deterministic samples cluster tightly near the reference tone curve, whereas random samples exhibit higher variance, particularly in low-to-mid intensity regions.

{Beyond output quality, Fig.~\ref{pimpim1} also summarizes hardware efficiency using an energy macro-model derived from Table~\ref{tab:delay_power_comparison}. 
With $C_{\text{total}}=1024$ memory columns, PIM-PIM processes $P=\lfloor C_{\text{total}}/N\rfloor=4$ pixels per batch and convert bit-sequential execution to bit-parallel LIM operation, providing an average throughput speedup of $53.9\times$ (642.5~FPS vs. 11.9~FPS at 200~MHz). Under the same model, the $sigmoid$ function exhibits comparable energy per frame for parallel and serial execution ($1.195\times10^{-5}$~J vs. $1.087\times10^{-5}$~J), while $\tanh$ favors the parallel design ($1.211\times10^{-5}$~J vs. $1.528\times10^{-5}$~J). Despite higher instantaneous power in the parallel mode, the substantial reduction in execution latency yields substantial improvements in Energy-Delay Product (e.g., $\sim$49$\times$ for $sigmoid$ and $\sim$68$\times$ for $\tanh$).}

\section{Conclusion}
\label{conc}
This work presented an MTJ-based, in-memory stochastic computing (SC) approach that mitigates the long-standing bit-serial, $O(N)$-latency bottleneck of conventional serial SC by converting stochastic streams into spatial ``bit bundles” and executing stochastic logic in parallel across the memory array. The proposed architecture tightly integrates data storage, deterministic in-memory bit-stream generation, parallel stochastic arithmetic for both basic arithmetic and complex transcendental functions, and optional parallel bit-stream-to-binary conversion within a unified LIM-enabled MTJ fabric. Circuit- and application-level evaluations show that the proposed array-parallel execution model sustains low-latency function evaluation while maintaining good accuracy in a representative image-processing pipeline, PIM-PIM, demonstrating that exploiting memory-level parallelism is a practical path to scalable, throughput-oriented in-memory SC.


\section*{Acknowledgment}
This work is supported by NSF grants 2019511, 2339701, 2609436, NASA grant 80NSSC25C0335, Vernon $\&$ Ruby Langlinais Non-Endowed Research Fund, Lockheed Martin Corporation Endowed Professorship Fund, and gifts from NVIDIA and Google.

\section*{Appendix}
\label{appen}
This appendix details the formulation of transcendental functions using truncated Maclaurin-series expansions, providing a hardware-friendly foundation for their realization in SC.
\newline
\ding{172} $\mathbf{sin(x)}$. 
The \textit{sine} function is a prominent transcendental 
function 
widely used in various applications. 
Implementing the \textit{sine} function in hardware presents significant challenges. 
The 7\textsuperscript{th} order truncated Maclaurin series expansion of $\mathbf{sin(x)}$ is given by:
{
\begin{equation}
\begin{aligned}
\label{sin_eq}
    \sin(x)&  =   \sum_{n=0}^{\infty}  \frac{( - 1 )^{n}}{( 2n + 1 )!}x^{ 2n + 1} \approx x  -  \frac{x^{3}}{3!}  +  \frac{x^{5}}{5!}  -  \frac{x^{7}}{7!} \\
           &\approx x  \left(  1  -  \frac{x^{2}}{6}   \left(   1  -  \frac{x^{2}}{20}   \left(   1  -  \frac{x^{2}}{42}  \right)  \right)  \right).
\end{aligned}    
\end{equation}
}

A basic design of this function can be implemented by cascading a chain of \texttt{AND} and \texttt{NAND} gates alongside a couple of \texttt{D-FF}s 
to achieve the desired level of decorrelation. 
An alternative design 
is based on a cascaded \texttt{NAND}-\texttt{AND} structure~\cite{9444648}. While pseudo-random sequences have been used in the 
SOTA designs~\cite{Trig-Parhi}, our proposed IMC architecture offers an efficient implementation by employing quasi-random patterns.

\noindent
\ding{173} $\mathbf{cos(x)}$.
The 8\textsuperscript{th} order truncated Maclaurin series expansion of $\mathbf{cos(x)}$ is given by:
{
\begin{equation}
\begin{aligned}
\label{cos_eq}
\cos(x)&  =   \sum_{n=0}^{\infty}  \frac{( - 1 )^{n}}{(2n)!} x^{ 2n} \approx  1  -  \frac{x^{2}}{2!}  +  \frac{x^{4}}{4!}  -  \frac{x^{6}}{6!} + \frac{x^8}{8!}\\ 
       &\approx  1  -  \frac{x^{2}}{2}   \left (   1  -  \frac{x^{2}}{12}   \left (   1  -  \frac{x^{2}}{30}   \left (   1  -  \frac{x^{2}}{56}   \right )    \right )    \right ).
\end{aligned}    
\end{equation}
}

\noindent Compared to the $\mathbf{sin(x)}$ design, the 
number of delay elements required in the input and intermediate stages differs for 
$\mathbf{cos(x)}$, due to variations in the polynomial structure and coefficient placement within the factorized form.

\noindent
\ding{174} $\mathbf{tanh(x)}$.
One widely used function in mathematical modeling and learning systems is the \textit{hyperbolic tangent} function~\cite{8610326}. The 7\textsuperscript{th} order truncated Maclaurin series expansion of $\mathbf{tanh(x)}$ is given by:
{
\begin{equation}
    \begin{aligned}
    \label{tanh_eq}
        \tanh(x)& \approx  x  -  \frac{x^{3}}{3}  +  \frac{2x^{5}}{15}  -  \frac{17x^{7}}{315}\\
                &\approx x  \left (   1  -  \frac{x^{2}}{3}  \left (   1  -  \frac{2x^{2}}{5}  \left (   1  -  \frac{17x^{2}}{42}   \right )   \right )   \right). 
    \end{aligned}
\end{equation}
}

The overall architecture for implementing this function closely follows the design of the  $\mathbf{sin(x)}$ function, with the primary difference being the use of distinct coefficient values in the polynomial expansion.

\noindent
\ding{175} $\mathbf{arctan(x)}$.
\label{atanx}
To derive the polynomial approximation for the \textit{arctan} (i.e., \textit{inverse tangent}) function in a from analogous to the Maclaurin series expansion, we begin with the infinite series expansion of {$\sum_{i=0}^{\infty}x^{i}$} as follows:
\vspace{-0.5em}
{
\begin{equation}
    \label{sum_inf}
    \sum_{i=0}^{\infty }x^{i}=\frac{1}{1-x},
\end{equation}
}

\noindent where $|x| < 1$. The $\mathbf{arctan(x)}$ formula is derived by the integral of the function $\frac{1}{1+x^{2}}$ as follows:
{
\begin{equation}
\label{atan_eq}
    \arctan(x) = \int \frac{1}{1+x^{2}}dx.
\end{equation}}
Hence, following the equation~(\ref{sum_inf}) we have:
{
\begin{equation}
    \label{atan_expansion}
    \frac{1}{1+x^{2}}=\frac{1}{1-(-x^{2})}=1-x^{2}+x^{4}-x^{6}+x^{8}-\cdots
\end{equation}
}

By integrating 
both sides of equation~(\ref{atan_expansion}) we have:
{
\begin{equation}
\label{atan_final}
\begin{aligned}
    \arctan(x) & =  x  -  \frac{x^{3}}{3}  +  \frac{x^{5}}{5}  -  \frac{x^{7}}{7}\\
               &\approx x  \left (   1  -  \frac{x^{2}}{3}   \left (   1  -  \frac{3x^{2}}{5}  \left (   1  -  \frac{5x^{2}}{7}   \right )   \right )  \right ).
\end{aligned}
\end{equation}
}

This design also mirrors the structure used for implementing the $\mathbf{sin(x)}$ function, with the key distinction being the use of different coefficient values. Such a modular representation enables a universal approach to designing transcendental 
functions within the same architectural framework, requiring only coefficient adjustments tailored to each specific function. 
\newline
\ding{176} $\mathbf{sigmoid(x)}$.
The \textit{sigmoid} function is a widely used transcendental 
activation function, particularly prominent in learning systems and neural networks.
The 5\textsuperscript{th} order truncated Maclaurin series expansion of the $\mathbf{sigmoid(x)}$ function is given by:
{
\begin{equation}
    \begin{aligned}
    \label{sigmoid_eq}
        sigmoid(x)\!&\! \approx  \frac{1}{2}  +  \frac{x}{4}  -  \frac{x^{3}}{48}  +  \frac{x^{5}}{480}\\
                  \!&\!\approx 1  -  \frac{1}{2}  \left (   1  -  \frac{x}{2}  \left (   1  -  \frac{x^{2}}{12}  \left (   1  -  \frac{x^{2}}{10}   \right )   \right )   \right ).
    \end{aligned}
\end{equation}
}


The design of this function closely resembles that of the $\mathbf{cos(x)}$ function, differing primarily in the arrangement of the delay elements in the intermediate stages and the specific coefficient values. By leveraging the same modular and universal design principles—-modifying only the coefficients within the SC module—-the architecture remains streamlined and reusable across functions. 
\newline
\ding{177} $\mathbf{Sinc( x )} = \frac{\mathbf{sin( x )}}{\mathbf{x}}$.
The \textit{Sinc} function is 
a fundamental component in signal processing and communication systems.
By dividing 
both sides of Equation~(\ref{sin_eq}) by $x$ ($x  \neq   0$), the 6\textsuperscript{th} order truncated Maclaurin series expansion of the $\mathbf{Sinc(x)}$ function can be derived as:
{
\begin{equation}
    \begin{aligned}
    \label{Sinc_eq}
        Sinc( x ) &= \frac{sin( x )}{x} \approx  1  -  \frac{x^{2}}{3!}  +  \frac{x^{4}}{5!}  -  \frac{x^{6}}{7!}\\
                  &\approx 1  -  \frac{x^{2}}{6}  \left (   1  -  \frac{x^{2}}{20}  \left (   1  -  \frac{x^{2}}{42}   \right )   \right ).
    \end{aligned}
\end{equation}
}

The design of this function closely follows that of the $\mathbf{sin(x)}$ function, with the primary distinction being the omission of the final stage in the polynomial expansion.
\newline
\ding{178} $\mathbf{e^{-x}}$.
The function $\mathbf{e^{-x}}$, commonly referred to as the decay function, is a fundamental 
component in probabilistic modeling and statistical analysis. The 5\textsuperscript{th} order truncated Maclaurin series expansion of the $\mathbf{e\textsuperscript{$-x$}}$ function is given by:
{
\begin{equation}
    \begin{aligned}
    \label{exp_eq}
        e^{ - x}&  =   \sum_{n=0}^{\infty}  \frac{( - x )^{ n }}{n!}  \approx  1  -  x  +  \frac{x^{ 2 }}{2!}  -  \frac{x^{ 3 }}{3!}  +  \frac{x^{ 4 }}{4!}  -  \frac{x^{ 5 }}{5!}\\
                &\approx  1  -  x  \left (   1  -  \frac{x}{2}  \left (  1-  \frac{x}{3}  \left (   1  -  \frac{x}{4}  \left (   1  -  \frac{x}{5}   \right )   \right )   \right )   \right ).
    \end{aligned}
\end{equation}
}

This function can be implemented by a series of \texttt{NAND} gates, each accompanied by a single delay element.
\newline
\ding{179} $\mathbf{ln(1+x)}$.
The natural \textit{logarithm} is widely used in machine learning and data analytics-—for instance, in computing entropy in decision trees, preprocessing skewed data, and normalizing or scaling feature values. 
The 5\textsuperscript{th} order truncated Maclaurin series expansion of the $\mathbf{ln(1+x)}$ function is as follows:
{
\begin{equation}
    \begin{aligned}
    \label{ln_eq}
        \ln( 1  +  x )  &=   \sum_{n=0}^{\infty}  \frac{( - x )^{ n }}{n}  \approx  x  -  \frac{x^{ 2 }}{2}  +  \frac{x^{ 3 }}{3}  -  \frac{x^{ 4 }}{4}  +  \frac{x^{ 5 }}{5}\\
                        &\approx  x \! \left ( \!  1  -  \frac{x}{2}  \left ( \!  1  - \! \frac{2 x}{3} \! \left ( \!  1  - \! \frac{3 x}{4} \! \left ( \!  1  -  \frac{4 x}{5} \!  \right ) \!  \right ) \!  \right ) \!  \right ).
    \end{aligned}
\end{equation}
}

This function can also be implemented similarly to the $\mathbf{e\textsuperscript{$-x$}}$ function, 
except the coefficient values are different.

\color{black}

\bibliographystyle{ACM-Reference-Format}
\bibliography{hassan, references}

\end{document}